\documentclass[aps,prd,twocolumn,showpacs,preprintnumbers,amsmath,amssymb,longbibliography]{revtex4-1}
\usepackage{hyperref}
\usepackage{bm,bbm}
\usepackage{graphics}
\usepackage{braket}
\usepackage{mathtools}
\usepackage{mathrsfs}
\usepackage{comment}
\usepackage{autobreak}
\usepackage{color}

\usepackage{tikz}
\usetikzlibrary{arrows,arrows.meta,intersections, calc,positioning,decorations.pathreplacing,decorations.pathmorphing,shapes}
\usetikzlibrary{patterns}
\usetikzlibrary{decorations.markings}
\usetikzlibrary{knots}

\let\Re\relax
\DeclareMathOperator{\Re}{Re}

\begin{document}

\title{Semi-classical saddles of three-dimensional gravity via holography}
\preprint{YITP-24-24}

\author{Heng-Yu Chen,$^{a,b}$ Yasuaki Hikida,$^b$ Yusuke Taki$^b$ and Takahiro Uetoko$^c$}

\affiliation{$^a$Department of Physics, National Taiwan University, Taipei 10617, Taiwan}
\affiliation{$^b$Center for Gravitational Physics and Quantum Information, Yukawa Institute for Theoretical Physics, Kyoto University, Kyoto 606-8502, Japan}
\affiliation{$^c$Department of General Education, National Institute of Technology, Kagawa College, Chokushicho 355, Takamatsu, Kagawa 761-8058, Japan}
%\date{\today}

\begin{abstract}

We find out the complex geometries corresponding to the semi-classical saddles of three-dimensional quantum gravity by making use of the known results of dual conformal field theory (CFT), which is effectively given by Liouville field theory. We examine both the cases with positive and negative cosmological constants. We determine the set of semi-classical saddles to choose from the homotopy argument in the Chern-Simons formulation combined with CFT results and provide strong supports from the mini-superspace approach to the quantum gravity. For the case of positive cosmological constant, partial results were already obtained in our previous works, and they are consistent with the current ones. For the case of negative cosmological constant, we identify the geometry corresponding a semi-classical saddle with three-dimensional Euclidean anti-de Sitter space dressed with imaginary radius three-dimensional spheres. 
The geometry is generically unphysical, but  the fact itself should not lead to any problems as derived from consistent dual CFT. We thus find an intriguing example, where the gravity path integral is performed over unphysical geometries.

\end{abstract}

\maketitle

%%%%%%%%%%%%%%%%%%%%%%%%%%%%%%%%%%%%%%%%%%%%%%%%%%%%%%%%%%%%%%%%%%%%%%

\section{Introduction}

One of the important issues in the path integral formulation of quantum gravity is to determine which geometries should be integrated over or in other words which saddle points should be summed over. 
We attack this problem by making use of holographic principle \cite{Susskind:1994vu}, which states that a gravity theory can be described by a lower dimensional non-gravitational particle theory. A famous example is given by the AdS/CFT correspondence \cite{Maldacena:1997re,Gubser:1998bc,Witten:1998qj}, where gravity theory on anti-de Sitter (AdS) space is dual to a lower dimensional conformal field theory (CFT). 
There are other examples like dS/CFT correspondence \cite{Witten:2001kn,Strominger:2001pn,Maldacena:2002vr}, where gravity theory on de Sitter (dS) space can be described by a lower dimensional CFT. The version of holography is less understood compared to the AdS/CFT correspondence, but there have been steady progresses like proposals on concrete examples, e.g.,  \cite{Anninos:2011ui,Hikida:2021ese,Hikida:2022ltr}. Applying the two types of holography, we study the semi-classical behaviors of gravity theory with negative or positive cosmological constant.

In order to make path integral convergent, we usually perform a Wick rotation from Lorentzian background geometry to Euclidean one. Similarly, it might be useful to integrate over complex geometries for the path integral of gravity theory. A canonical example of complex geometry is given by the no-boundary proposal, which states that our universe starts from nothing \cite{Hartle:1983ai}. More concretely speaking, the universe is made by gluing a hemi-sphere and a half of dS space. Recently, it was analyzed which complex geometry is allowed for quantum gravity in \cite{Witten:2021nzp} by applying a criterion proposed by \cite{Louko:1995jw,Kontsevich:2021dmb}. A naive guess is that the gravity path integral can be given by the sum of the contributions from ``allowable'' complex geometries dressed by perturbative corrections. It is natural to think that possible semi-classical geometry realized by a path integral saddle can be read off from dual CFT. We would like to examine this from a concrete example and compare with the criterion of allowable complex geometry.

If we want to study quantum aspects of gravity theory, then we need detailed information on dual CFT. Generically, supersymmetric theory provides us some exact results, but it is notoriously difficult for de Sitter space to preserve any supersymmetry. In this paper, we thus decide to examine three-dimensional gravity since its dual two dimensional CFT has a infinite dimensional symmetry, which enables us to obtain exact answers. A concrete proposal of dS/CFT correspondence was provided in \cite{Hikida:2021ese,Hikida:2022ltr} and developed further in \cite{Chen:2022ozy,Chen:2022xse}. In fact, the semi-classical limit of dS gravity was examined in \cite{Chen:2023sry,Chen:2023prz} in this way. This work is a continuation of the work by extending the analysis to AdS gravity.

As the effective CFT dual to three-dimensional gravity, we utilized Liouville field theory. In fact, exact two- and three-point functions were obtained in \cite{Dorn:1994xn,Zamolodchikov:1995aa} and their ``semi-classical'' expressions were analyzed in \cite{Zamolodchikov:1995aa,Harlow:2011ny}. The use of Liouville field theory may be justified as follows. It was proposed in \cite{Gaberdiel:2010pz} that higher spin gravity theory on AdS$_3$ by \cite{Prokushkin:1998bq} is dual to the W$_N$ minimal model, which has a coset description
\begin{align}
\frac{SU(N)_k \times SU(N)_1}{SU(N)_{k+1}} \, , 
\end{align}
where the central charge is 
\begin{align}
c = (N-1) \left(1 - \frac{N (N+1)}{(N+k) (N+k+1) } \right) \, .
\end{align}
The gravity theory includes gauge fields with higher spin $s=2,3,\cdots$, which can be organized by Chern-Simons gauge theory with an infinite dimensional algebra $\mathfrak{hs}[\lambda]$. The parameters in the gravity theory are given by the Newton constant $G$ and the AdS length $\ell_\text{AdS}$. The central charge of dual CFT is known to be related to them as \cite{Brown:1986nw}
\begin{align}
c = \frac{3 \ell_\text{AdS}}{2 G} \, .
\end{align}
Thus the small $G$ limit corresponds to the large $c$ limit with fixed $\lambda = N/p$. A different version of the duality was argued in \cite{Castro:2011iw,Gaberdiel:2012ku,Perlmutter:2012ds}.  In the gravity theory, we truncate the higher spin algebra $\mathfrak{hs}[\lambda]$ to $\mathfrak{sl}[N]$ by setting $\lambda =  - N$. The gravity theory includes also matter fields, whose effects can be neglected at the limit of $G \to 0$. In this letter, we focus on the case with $N=2$ and describe the pure gravity by $\mathfrak{sl}(2)$ Chern-Simons theory \cite{Achucarro:1986uwr,Witten:1988hc}. The dual CFT is proposed to be given by the same coset but with $N=2$ and large central charge $c$. This leads to a negative real $k \to - N - 1$, which implies the CFT is non-unitary. It was shown in \cite{Creutzig:2021ykz} that the correlation functions of the non-unitary CFT can be computed by Liouville field theory. In this sense, the semi-classical limit of pure gravity on AdS$_3$ can be captured by Liouville field theory. Following the prescription of \cite{Maldacena:2002vr}, we analytically continue AdS gravity to dS gravity by changing $\ell_\text{AdS}$ to $ - i \ell_\text{dS}$ along with a proper change of coordinates. This implies that the dual CFT has an imaginary central charge \cite{Strominger:2001pn,Ouyang:2011fs}
\begin{align}
c = - i c^{(g)} = - i\frac{ 3 \ell_\text{dS}}{2 G}
\end{align}
with large real $c^{(g)}$, and its effective description is given by Liouville field theory as well.

\section{Semi-classical saddles in dS gravity}

\label{sec:2ptresurgence}

We start by reviewing the analysis in \cite{Chen:2023prz,Chen:2023sry}, where the semi-classical saddles in gravity theory on dS$_3$ were derived by making use of the exact results of Liouville field theory. We are interested in the wave functional of universe used for the no-boundary proposal by Hartle and Hawking. 
In terms of path integral, the wave functional can be written as
\begin{align} \label{eq:wfdS}
\Psi_\text{dS} [h] = \int \mathcal{D} g  e^{- I[g]  } \, , \quad
\end{align}
where the action is
\begin{align} \label{eq:action}
I [g] = I_\text{EH} + I_{\text{GH}} + I_\text{CT} \, .
\end{align}
The Einstein-Hilbert action on the base manifold $\mathcal{M}$ and the Gibbons-Hawking  boundary terms on its boundary $\partial \mathcal{M} $ are given by
\begin{align}
\begin{aligned} \label{eq:action2}
& I_\text{EH} = - \frac{1}{16 \pi G} \int_\mathcal{M} d^3 x \sqrt{g} (R - 2 \Lambda) \, , \\
&I_\text{GH} + I_\text{CT} = \frac{1}{8 \pi G} \int_{\partial \mathcal{M}} d^2 x \sqrt{h} (K + \sqrt{ - \Lambda}) \, ,
\end{aligned}
\end{align}
where $K$ is the extrinsic curvature and $I_\text{CT}$ includes a counter term.
The cosmological constant is set as $\Lambda = \ell_\text{dS}^{-2}$.
The integration over the metric $g$ is performed under the boundary condition $g = h$ at the future infinity. It was proposed in \cite{Maldacena:2002vr} that the wave functional can be computed by the dual CFT partition function as
\begin{align}
  \Psi_\text{dS} [h]  = Z_\text{CFT} [h] \, .
\end{align}
Assuming that the path integral over $g$ is dominated by saddle points $g = g_{(n)}$ with integer label $n$ for small $G$, the wave functional can be decomposed by
\begin{align}
\Psi_\text{dS} [h] = \sum_n \Psi_n  [h] \, , \quad \Psi_n [h] \sim e^{- I [g_{(n)}]} \, .
\end{align}
Note that $ e^{- I [g_{(n)}]} $ is regarded as a non-perturbative contribution as $I[g_{(n)}]$ is proportional $1/G$, which is dressed by perturbative corrections in $G$. 
In the following we determine the set of saddle points in the gravity path integral from the dual CFT, i.e., Liouville field theory.

The action of Liouville field theory is given by
\begin{align}
S_\text{L} [\phi] = \frac{1}{\pi} \int d^2 z \left[ \partial \phi \bar \partial \phi + \frac{Q}{4} \sqrt{g} \mathcal{R} \phi + \pi \mu e^{2 b \phi}\right] \, .
\end{align}
Here $g_{ij}$ is the worldsheet metric, $g = \det g_{ij}$ and $ \mathcal{R}$ is the scalar curvature. We usually set $g_{ij} = \delta_{ij}$ and introduce complex coordinates $z , \bar z $. 
The central charge $c$
and the background charge $Q$ are given by
\begin{align}
c = 1 + 6 Q^2 \, , \quad Q = b + b^{-1} \, . \label{eq:Lcentral}
\end{align}
We realize the large central charge limit by $b \to 0$. We are interested in the correlation functions of the form
\begin{align}
\begin{aligned}
 \left \langle \prod_{j=1}^m e^{2 \alpha_j \phi (z_j , \bar z_j)} \right \rangle &= \int \mathcal{D} \phi e^{-  S_\text{L}[\phi]} \prod_{j=1}^m e^{2 \alpha_j \phi (z_j , \bar z_j) }  \\
& \equiv \int \mathcal{D} \phi e^{- \frac{1}{b^2} \tilde  S_\text{L}[\phi_c]} \, ,
\end{aligned}
\end{align}
where $\alpha_j$ behaves as $\alpha_j = \eta_j/b$ with fixed $\eta_j$. We restrict ourselves to the case with $0 < \eta_j < 1/2$, which is called as Seiberg bound \cite{Seiberg:1990eb}.
These operators are usually referred as heavy. At this limit, the insertions of vertex operators can be included as a part of action in $\tilde S_\text{L} [\phi_c]$, and $b^2$ can be regarded as the expansion parameter.
The equation of motion from the modified action is 
\begin{align}
\partial \bar \partial \phi_c  = 2 \lambda e^{\phi_c} -2 \pi \sum_i \eta_i \delta^2 (z -z_i) \, ,
\end{align}
where  $\phi_c = 2 b \phi$ and $\lambda = \pi \mu b^2$ is kept finite at the limit $b \to 0 $. We can easily see that
\begin{align}
\phi_{c(n)} = \phi_{c(0)} + 2 \pi i n
\end{align}
with integer $n$ can be a solution to the equation of motion as long as $\phi_{c(0)}$ is one of its solutions.  This means that there are saddle points labeled by an integer $n$ and the correlation functions can be decomposed into 
\begin{align}
 \left \langle \prod_{j=1}^m e^{\eta_j \phi_c (z_j , \bar z_j)} \right \rangle \sim \sum_n  e^{- \frac{1}{ b^{2}} \tilde  S_\text{L}[\phi_{c(n)}]} \, . \label{eq:Ldec}
\end{align}
The semi-classical saddles are dressed by perturbative quantum corrections in $b^2$, which are ignored here.

The small $b$ behaviors of correlation functions were examined in \cite{Harlow:2011ny} based on the exact results obtained in \cite{Dorn:1994xn,Zamolodchikov:1995aa}. In this letter, we consider specifically the two-point function of heavy operator with $0 < \eta < 1/2$, which is dual to the conical defect geometry with angle $2 \pi (1 - 2 \eta)$. For small $b$,
the two point function behaves as 
\begin{align} \label{eq:smallb}
& \left \langle e^{\eta \phi_c (z_1 , \bar z _1 )}  e^{\eta \phi_c (z_2 , \bar z _2 ) }  \right \rangle  \\
&\sim \delta (0) |z_{12}|^{- 4 \eta (1 - \eta) /b^2 } \left( e^{- \pi i (1 - 2 \eta)/b^2} - e^{\pi i (1 -2 \eta)/b^2} \right)^{\pm 1} \nonumber \\
& \quad \times
\exp \left\{  - \frac{2}{b^2} [ (1 - 2 \eta) \ln (1 - 2 \eta) - (1 - 2 \eta) ] \right\} \nonumber
\end{align}
with $+1$ for $\Re b^{-2} < 0$ and $-1$ for $\Re b^{-2} > 0$. 
The dS/CFT map suggests that the wave functional of the conical defect geometry is written in terms of the two-point function in Liouville field theory as
\begin{align}
\Psi_\text{dS} = \left \langle e^{\eta \phi_c (z_1 , \bar z _1 )}  e^{\eta \phi_c (z_2 , \bar z _2 ) }  \right \rangle \, .
\end{align}
From $c = 1 + 6 (b + b^{-1})^2$, we obtain
\begin{align}
\frac{1}{b^{2}} = - i \frac{c^{(g)}}{6} - \frac{13}{6} + \mathcal{O} ((c^{(g)})^{-1}) \, .
\end{align}
As in \cite{Hikida:2021ese,Hikida:2022ltr,Chen:2023prz,Chen:2023prz}, we fix $c^{(g)}$ to be real since $c^{(g)}$  is directly related to a physical quantity.
Therefore, we conclude that $\text{Re}\, b^{-2} < 0$ in the current case.
Decomposing the two-point function as in \eqref{eq:Ldec}, we have from \eqref{eq:smallb}
\begin{align} \label{eq:dSresult}
\Psi_\text{dS} \sim \sum_{n=-1,0} (-1)^n e^{S_\text{GH}^{(n)}/2 + i \mathcal{I} } \, , \quad 
S_\text{GH}^{(n)} = \frac{(2 n + 1) \pi \ell_\text{dS}}{2 G} \, . 
\end{align}
Here the real part, $S_\text{GH}^{(n)}$, is a contribution to the Gibbons-Hawking entropy \cite{Bekenstein:1973ur,Hawking:1975vcx,Gibbons:1976ue,Gibbons:1977mu}  from a semi-classical saddle point in $G$ labeled by $n$. 
Here and in the following we set $\eta = 0$ just for the simplicity of expression.
The sum is taken over $n=-1,0$.
The imaginary part, $\mathcal{I}$, is independent of saddle points.
The case with non-zero $\eta$ and the phase factor with $\mathcal{I}$ are analyzed in \cite{Chen:2024qmn}.

We would like to interpret the result  \eqref{eq:dSresult} purely from the gravity viewpoints.
As discussed in \cite{Witten:2021nzp}, we may consider the ansatz for the metric as
\begin{align} \label{eq:ansatzdS}
 ds^2 = \ell_\text{dS}^2 \left[ \left(\frac{d\theta (u)}{du}\right)^2 d u^2 + \cos ^2 \theta  (u) d \Sigma^2 \right]\, .
\end{align}
We set $d \Sigma^2$ as the metric of $\mathbb{S}^2$ and $\theta (u)$ as a holomorphic function of $u$.
We are interested in the universe starting from nothing and approaching to dS$_3$. We thus set $\theta =   (n + 1/2)\pi $ at $u=0$ and $\theta = i u$ for $u \to \infty$. This leads to a family of geometry labeled by $n$. In order to make the geometry as a standard solution of the Einstein equation, we may set $\theta = (n + 1/2) \pi  (1 - u)$ for $0 \leq u \leq 1$ and $\theta = i (u - 1 )$ for $1 < u $. 
The wave functional of universe is given by \eqref{eq:wfdS}, which can be decomposed as in \eqref{eq:dSresult}.
The real part, $S^{(n)}_\text{GH}$, can be evaluated from the Einstein-Hilbert action with the Euclidean geometry realized for $0 \leq u \leq 1$, and \eqref{eq:dSresult} 
can be reproduced. On the other hand, the phase factor $\exp (i \mathcal{I}^{(n)})$ comes from the Lorentzian region realized for $1 < u$. 
From the purely gravity analysis, we cannot determine which integer $n$ should be summed over.
Only with the help of CFT, we can see that the sum is taken over $n = -1,0$. The geometry is the same as the one used for the no-boundary proposal by Hartle and Hawking and consistent with the result obtained by the criterion of allowable complex geometry in \cite{Louko:1995jw,Kontsevich:2021dmb,Witten:2021nzp}.

\section{Extension to AdS gravity}

We then move to the AdS case, where the gravity partition function is written in terms of path integral as
\begin{align} \label{eq:AdSpath}
\mathcal{Z}_\text{AdS} [h] = \int \mathcal{D} g e^{- I [g]} \, . 
\end{align}
Here the action is \eqref{eq:wfdS} with \eqref{eq:action2}. The cosmological constant is set as $\Lambda = - \ell_\text{AdS}^{-2}$. The AdS boundary is located at the spatial infinity and the metric is set as $g = h$ at the boundary. The map of AdS correspondence is well known as GKP-Witten relation \cite{Gubser:1998bc,Witten:1998qj},
where the gravity partition function is computed by the CFT partition function as
\begin{align}
\mathcal{Z}_\text{AdS} [h] = Z_\text{CFT} [h] \, .
\end{align}
The path integral \eqref{eq:AdSpath} may be dominated by saddle points $g = g_{(n)}$ for small $G$, and the gravity partition function may be decomposed as
\begin{align}
\mathcal{Z}_\text{AdS} [h]  = \sum_n \mathcal{Z}_n [h] \, , \quad \mathcal{Z}_n [h] \sim e^{- I [g_{(n)}]} \, .
\end{align}
As in the dS case, we determine the saddle points of the gravity path integral from dual CFT.

We again consider two-point function of heavy operator in Liouville field theory, which is dual to the conical defect geometry with angle $2 \pi (1 - 2 \eta)$. The gravity partition function is related to the two-point function as
\begin{align}
\mathcal{Z}_\text{AdS} = \left \langle e^{\eta \phi_c (z_1 ,\bar z_1)} e^{\eta \phi_c (z_2 , \bar z_2)}\right \rangle \, .
\end{align}
In this case, we have $\text{Re} \, b^{-2} \sim \text{Re} \, c/6  > 0$.
From \eqref{eq:smallb}, we can decompose the partition function as 
\begin{align} \label{eq:AdSresult}
\mathcal{Z}_\text{AdS} \sim \sum_{n=0,1,2,\ldots}  \Theta_n \mathcal{Z}_0\, , \quad 
\Theta_n = e^{\frac{\ell_\text{AdS}}{2 G} n \pi i } \, ,
\end{align}
where we set $\eta = 0$ as before.
The phase factor $\Theta_n$ depends on the label $n$ of semi-classical saddle point in $G$ and the sum is taken over $n= 0,1, 2, \ldots$.
The case with non-zero $\eta$ and the $n$-independent part, $\mathcal{Z}_0$, will be analyzed in \cite{Chen:2024qmn} (see, e.g., \cite{Chang:2016ftb,Chandra:2022bqq,Abajian:2023bqv} for related works).

We would like to interpret the result from the purely gravity viewpoint. As in \eqref{eq:ansatzdS}, we assume the form of metric as
\begin{align} \label{eq:ansatzAdS}
 ds^2 = \ell_\text{AdS}^2 \left[ \left(\frac{d\theta (u)}{du}\right)^2 d u^2 + \sinh ^2 \theta  (u) d \Sigma^2 \right]\, .
\end{align}
As before, we set $d \Sigma^2$ as the metric of $\mathbb{S}^2$ and $\theta (u)$ as a holomorphic function of $u$. We consider the geometry which approaches to Euclidean AdS$_3$ as $u \to \infty$ and truncates at $u=0$. Thus we assign $\theta =  n \pi i $ with integer $n$ at $u=0$ and $\theta = u$ for $u \to \infty$. To make the geometry as a standard solution of Einstein equation, we may set $\theta =  n \pi i (1 - u)$ for $0 \leq u \leq 1$ and $\theta = (u - 1 )$ for $1 < u $. For $1 < u$, the geometry is Euclidean AdS$_3$, but for $0 \leq u \leq 1$, the geometry becomes imaginary radius $\mathbb{S}^3$ or with three Lorentzian time directions. Thus, we may conclude that the geometry labeled by non-zero $n$ is unphysical. 
In particular, it does not satisfy the criterion of allowable complex geometry by \cite{Louko:1995jw,Kontsevich:2021dmb,Witten:2021nzp}.

In order to confirm the validity of the above counter-intuitive conclusion, we examine the meaning of the integer $n$ in terms of gravity theory. 
For the purpose, it is convenient to work with the first order formulation, where the gravity action can be rewritten in terms of Chern-Simons action as \cite{Achucarro:1986uwr,Witten:1988hc}
\begin{align}
\begin{aligned}
&I_\text{EH} = k I_\text{CS} [A] - k I_\text{CS} [\tilde A] \, , \\
&I_\text{CS} [A] = \frac{1}{4 \pi} \int \text{tr} \left[ A \wedge dA + \frac{2}{3} A \wedge A \wedge A \right]
\end{aligned}
\end{align}
up to boundary contributions (see \cite{Chen:2024qmn} for details). The Chern-Simons level is related to the gravitational parameters as $k = - i \kappa = - i \ell_\text{dS}/(4G)$ or $k = \ell_\text{AdS}/(4G)$. 
Let us first consider the dS case. Then, as mentioned above, the contribution to the Gibbons-Hawking entropy comes from the Euclidean geometry. For examining only the region, we can set $A , \tilde A \in \mathfrak{su}(2)$, i.e., an algebra associated with a compact group. See, e.g., \cite{Witten:1989ip} and appendix A of \cite{Chen:2023sry}. The Chern-Simons action is invariant under a small gauge transformation. However, for generic level $k$, a large gauge transformation is not a symmetry of the action anymore. A large gauge transformation generates gauge configuration where the value of the action has extra integer contribution as
\begin{align}
I_\text{CS} \to I_\text{CS} + 2 \pi i \mathbb{Z} \, .
\end{align}
In fact, the Chern-Simons action counts how many times the gauge configuration wraps $\mathbb{S}^3$ (i.e., $\pi_3\left( \mathbb{S}^3\right) = \mathbb{Z}$, see \cite{Anninos:2020hfj,Hikida:2022ltr} for examples).
We can see that each geometry labeled by integer $n$ leads to the same Chern-Simons action.

In order to apply the argument to the AdS case, it is useful to recall how the hemi-sphere is connected to the half of dS$_3$. We can describe Lorentzian dS$_3$ and Euclidean dS$_3$ by a hypersurface in $\mathbb{R}^4$ with a flat metric;
\begin{align}
 \epsilon_0 (X^0)^2 + \epsilon_1 (X^1)^2 + \epsilon_2 (X^2)^2 + \epsilon_3 (X^3)^2 = \Lambda^{-1} \, ,\label{eq:hyds}
\end{align}
where $\Lambda^{-1} = \ell_\text{dS}^2$. Lorentzian dS$_3$ is described with $- \epsilon_0 = \epsilon_1 = \epsilon_2 = \epsilon_3 = 1$. This geometry is glued to Euclidean dS$_3$ (or $\mathbb{S}^3$) with replacing $\epsilon_0 = -1 $ by $\epsilon_0 =1$ at $X_0 = 0$. 
Let us come back to the AdS case. We start from Euclidean AdS$_3$ and we may analytically continue the Euclidean geometry to Lorentzian one. We describe Euclidean AdS$_3$ by a hypersurface \eqref{eq:hyds} 
with $\epsilon_0 = \epsilon_1 = \epsilon_2 = - \epsilon_3 = 1$ and $\Lambda^{-1} = - \ell_\text{AdS}^2$. As before, we may perform a Wick rotation by replacing $\epsilon_0 = 1$ by $\epsilon_0 = -1$, which leads to the Lorentzian AdS$_3$. The geometry cannot have non-trivial winding number, thus it does not lead to a non-trivial topological number. In order to have a non-trivial quantity, we perform a different Wick rotation by replacing $\epsilon_3 = - 1$ by $\epsilon_0 = 1$.
The geometry is unphysical as it defines $\mathbb{S}^3$ with imaginary radius $i \ell_{\text{AdS}}$. Even so, in terms of Chern-Simons gauge theory, there is no problem to construct such gauge configuration.

\section{Mini-superspace approach to quantum gravity}

As our conclusion is rather surprising, we would like to provide further argument supporting our claim.
For the purpose, we utilize the mini-superspace approach to the quantum gravity, see, e.g.,  \cite{Feldbrugge:2017kzv,Caputa:2018asc,Donnelly:2019pie,DiTucci:2020weq} for recent developments and \cite{Lehners:2023yrj} for a review. 
In order to confirm the validity of the approach, we first reproduce the previous result in \cite{Chen:2023prz,Chen:2023sry} for the dS case.
We use the ansatz for metric as
\begin{align}
    ds^2=\ell_\text{dS}^2\left[N(\tau)^2 d\tau^2+a(\tau)^2d \Sigma^2 \right]\, ,
\end{align}
which slightly generalizes \eqref{eq:ansatzdS}.
Without loss of generality, we can set $0 \leq \tau \leq 1$. We would like to evaluate the path integral \eqref{eq:wfdS}. The gauge redundancy allows us to fix $N(\tau)$ to be constant $N$, and the path integral can be reduced to \cite{Halliwell:1988wc}
\begin{align} \label{eq:wf2}
 \Psi_\text{dS} = \int_\mathcal{C} d N \int \mathcal D a (\tau) e^{- I[a;N]  - I_{\text{CT}}}
\end{align}
with
\begin{align} \label{eq:IaN}
 I[a;N] = -\frac{\ell_{\text{dS}}}{2G}\int_0^1d\tau \,N\left[\frac{1}{N^2} \left(\frac{d a}{d \tau}\right)^2-a^2+1\right] \, .
\end{align}
The contour over $N$ is denoted by $\mathcal{C}$.
Here we set the boundary conditions
$
a(0)=0  , a(1)=a_1
$
and $I_{\text{CT}}$ cancels the divergence proportional to $a_1^2$.

We first integrate $a(\tau)$ out. For this, we decompose $a(\tau) = \bar a^{(N)} (\tau) + A (\tau)$, where $\bar a^{(N)} (\tau)$ is a solution to the equation of motion for the action \eqref{eq:IaN} and $A(\tau)$ represent small fluctuations around it. A solution to the equation of motion 
$
    d^2a / d\tau^2 + N^2 a=0
$
is given by 
\begin{align}
    \bar{a}^{(N)}(\tau)=\frac{a_1}{\sin N}\sin\left(N\tau\right)\, .
\end{align}
Here we have imposed the boundary conditions 
$
a(0)=0  , a(1)=a_1
$.
After integrating the fluctuations $A(\tau)$, the wave functional \eqref{eq:wf2} becomes (see \cite{Chen:2024qmn} for derivation)
\begin{align} \label{eq:wf3}
 \Psi = \int_\mathcal{C} d N  \left( \frac{1}{\sqrt{N}\sin N} \right)^{\frac12} e^{ \frac{\ell_\text{dS}}{2G} (N + a_1^2 \cot N )  - I_{\text{CT}} }
\end{align}
up to an irrelevant overall normalization.

\begin{figure}
\centering
\includegraphics[width=0.45\linewidth]{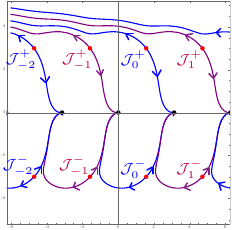}
\hspace{1.5mm}
\includegraphics[width=0.45\linewidth]{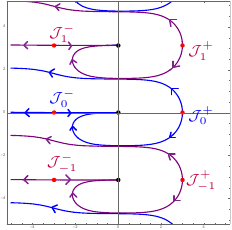}
    \caption{Saddle points are given by red points and steepest descent paths are drawn by solid lines. Black dots denote the singularities of the integrand. Right and left figures are for $\Lambda = (\ell_\text{dS} + i \epsilon)^{-2}$ and $\Lambda = - (\ell_\text{AdS} + i \epsilon)^{-2}$, respectively. }
    \label{fig:lshift}
\end{figure}

Now the problem is to find out a proper contour $\mathcal{C}$, for the integration over $N$. For this, we first find out the saddle points by solving $\partial I[\bar a^{(N)};N]/\partial N = 0$. The solutions are 
\begin{align}
\begin{aligned}
    N^+_n&=\left(n+\frac{1}{2}\right)\pi+i\ln\left(a_1+\sqrt{a_1^2-1}\right)\,,\\
    N^-_n&= \left(n+\frac{1}{2}\right)\pi-i\ln\left(a_1+\sqrt{a_1^2-1}\right)\,,
    \end{aligned}
\end{align}
with $n \in \mathbb{Z}$, which are represented by red points in the left panel of fig.\ \ref{fig:lshift}.
We next look for the paths of steepest descent from the saddles. We find a subtlety that some of lines for the steepest descent and steepest ascent coincide with each other. In order to avoid this, we introduce a regulator as $\ell_\text{dS} \to \ell_\text{dS} + i \epsilon$ with $\epsilon > 0$
\footnote{The necessity of regulator was discussed in \cite{Feldbrugge:2017kzv}, see also \cite{Honda:2024aro}.
The sign of $\epsilon$ was chosen to be consistent with the CFT analysis \cite{Chen:2024qmn}.}.
The contribution of each saddle to the wave functional \eqref{eq:wf3} are
\begin{align}
\Psi_n^\pm \sim e^{\frac{ (n +1/2) \ell_\text{dS} \pi}{2 G} } (2 a_1)^{ \pm \frac{i\ell_\text{dS} + \epsilon}{2G}} 
\end{align}
for large $a_1$. The contributions from the saddles $N^-_n$ are suppressed for large $a_1$, and only the contributions from the saddles $N^+_n$ are relevant.
The paths of steepest descent are written down as solid lines in fig.\ \ref{fig:lshift}. We denote that the path through $N^\pm_n$ by $\mathcal{J}^\pm_n$ and the orientations of paths are defined to be positive in the positive directions of the real and imaginary axes.

We may choose the integration contour for $N$ to be $i \mathbb{R}_+$ to naively preserve future time direction. 
However, this contour does not reproduce the result in \eqref{eq:dSresult} obtained from dual CFT, so we need to find another contour. 
There are infinitely many candidates for such a contour. 
Among them, we find that the proper contour is 
\begin{align}
\mathcal{C} = -\mathcal{J}^+_{0} + \mathcal{J}^-_{0} + \mathcal{J}^+_{-1} \, .
\end{align}
The geometry described by the saddle points looks complicated. However, as in \cite{Lehners:2021mah}, by Cauchy's theorem, we introduce a time coordinate as:
\begin{align}\label{Ttau}
T(\tau) =  - \left(n + \frac{1}{2} \right) \pi (1 - \tau)^q + i \ln (2 a_1 ) \tau^q \, .
\end{align}
This encodes a deformation of the contour interpolating between $\tau=0$ and $\tau=1$, along which the geometric criterion of \cite{Kontsevich:2021dmb} can be satisfied.
For $q \to \infty$, the geometry approaches to the one discussed below \eqref{eq:ansatzdS}.

We then move to the main problem with negative cosmological constant. As in the previous case, we consider the ansatz as
\begin{align}\label{eq:minisuperspace}
ds^2=\ell_\text{AdS}^2\left[N(r)^2dr^2+a(r)^2d\Sigma^2\right] 
\end{align}
with $d\Sigma^2$ being the metric of $\mathbb{S}^2$.
In this case, $r$ is the radial coordinate taking a value in $0<r<1$. As in \eqref{eq:wf2}, the partition function evaluated by integrating over the metric can be reduced to 
\begin{align}\label{pfmini}
    \mathcal{Z}=\int_{\mathcal{C}}dN\int \mathcal{D}a (r) e^{-I[a;N] - I_{\text{CT}}}\,,
\end{align}
where the action is expressed as
\begin{align}\label{actionmini}
    I[a;N]=-\frac{\ell_\text{AdS}}{2G }\int_0^1dr\, N\left(\frac{1}{N^2} \frac{d ^2 a}{dr^2}+a^2+1\right) \, .
\end{align}
We assign the Dirichlet boundary conditions $a(0) = 0 $ and $a(1) = a_1$.

We first integrate over $a(r) = \bar a^{(N)}(r) + A(r)$, where $a^{(N)}(r)$ is the saddle point and $A(r)$ are fluctuations around it. The equation of motion for $a(r)$ is given by 
$
    d^2a /d\tau^2 - N^2 a=0
$
and the solution subject to the Dirichlet boundary conditions is
\begin{align}\label{eq:DDsolution}
    \bar{a}^{(N)}(r)=\frac{a_1}{\sinh N}\sinh (Nr)\,.
\end{align}
We further integrate out the fluctuations, $A(r)$.
The partition function is now written as (see \cite{Chen:2024qmn} for derivation)
\begin{align} \label{eq:pf}
 \mathcal{Z} = \int_\mathcal{C} d N  \left( \frac{1}{\sqrt{N}\sinh N} \right)^{\frac12} e^{\frac{\ell_\text{dS}}{2G} (N + a_1^2 \coth N )  - I_{\text{CT}} } 
\end{align}
up to an overall factor.

We then determine the contour $\mathcal{C}$.
Solutions to $\partial I[\bar{a}^{(N)};N]/\partial N=0$ are listed as
\begin{align}
\begin{aligned}
    &N^+_n=  n \pi i  + \ln \left( a_1 + \sqrt{a_1^2 + 1}  \right) \, , \\
    &N^-_n=  n \pi i  - \ln \left( a_1 + \sqrt{a_1^2 + 1}  \right) \, , 
\end{aligned}
\end{align}
with $n \in \mathbb{Z}$. The steepest descent lines originating these saddle points can be obtained. However, as in the previous case, some of lines for the steepest descent and ascent coincide with each other, which may require a regularization as, say, $\ell_{\text{AdS}}\to\ell_{\text{AdS}} + i\epsilon$. The steepest descent paths after the shift are depicted in the right panel of fig.\ \ref{fig:lshift}.
We denote that the steepest descent path through $N^\pm_n$ by $\mathcal{J}^\pm_n$ and its orientations are defined to be positive in the positive directions of the real and imaginary axes.

In this case, it is natural to integrate over $N$ along the positive real line. 
This integration contour can be deformed into
\begin{align}
\mathcal{C} = 
\sum_{n=0}^{\infty}\mathcal{J}^+_n -\sum_{n=1}^{\infty}\mathcal{J}^-_n \, .
\end{align}
We thus take the saddle points $N^+_n$ with $n=0,1,2\ldots$ and $N^-_n$ with $n=1,2,\ldots$.
Each contribution from the saddle point to the partition function is
\begin{align}
    \mathcal{Z}^\pm_n \sim e^{\frac{n \pi i (\ell_\text{AdS} + i \epsilon)}{2G}}(2a_1)^{\pm \frac{\ell_\text{AdS}}{2G}}
\end{align}
for large $a_1$. 
As before, we can neglect the contribution $\mathcal{Z}^-_n$ for large $a_1$. 
Thus the semi-classical limit of partition function is expressed as the convergent sum of $\mathcal{Z}^+_n$ with $n=0,1,2,\ldots$, which actually reproduces the result in \eqref{eq:AdSresult} obtained from the dual CFT.
Therefore, we conclude that the natural contour we took is the proper contour. 
As in the previous dS case, we introduce a new radial coordinate:
\begin{align}
R(t) = - n \pi i (1 - r)^q + \ln (2 a_1) r^q \, .
\end{align}
Again the geometry approaches to the one discussed below \eqref{eq:ansatzAdS} for $q \to \infty$.

\section{Discussion}

In this letter, we proposed the complex geometries corresponding to the semi-classical saddles in three-dimensional quantum gravity with the help of dual CFT. 
In particular, for the case of negative cosmological constant, the geometry is claimed to be given by Euclidean AdS$_3$ attached with imaginary radius $\mathbb{S}^3$'s. The geometry should be unphysical, however this fact itself should not leads to any contradiction as it was derived from a consistent dual CFT. For instance, the partition function is real after summing over the geometry as
\begin{align}
 - i e^{\pi i \frac{\ell_\text{AdS}} {4G}} \sum_{n=0}^\infty e^{ n  \pi i \frac{\ell_\text{AdS}} {2G}} = \frac{1}{2 \sin \pi \frac{\ell_\text{AdS} }{4G} }
\end{align}
up to an overall phase factor.
It would be also interesting to investigate the relation with ``time-like entanglement entropy'' \cite{Doi:2022iyj,Doi:2023zaf}, which seems to capture the information of the similar attached imaginary geometry.

We would like to relate the result to the arguments based on the criterion of \cite{Louko:1995jw,Kontsevich:2021dmb,Witten:2021nzp}.
The criterion is that, in $D$-dimensional complex geometry, the path integral over real $p$-form field for $p=0,1,\ldots,D$ with their free action should converge. 
If we choose a gauge of diagonal metric as $g_{ij} = \text{diag} (\lambda_1 , \lambda_2 , \ldots , \lambda_D )$, then the criterion can be simplified as
\begin{align} \label{eq:criterion}
 \sum_{i=1}^D |\text{arg} \lambda_i| < \pi \, .
\end{align}
It is quite sensible as it requires that the maximal number of time-like directions is one. However, as argued in \cite{Witten:2021nzp}, the condition is neither necessary nor sufficient. 
Among the examples mentioned in \cite{Witten:2021nzp}, the geometry for no-boundary proposal and Euclidean Kerr black hole with asymptotic AdS satisfy the criterion (see \cite{Basile:2023ycy} for an analysis in the current setup), while Euclidean Kerr black hole with asymptotic flat space-time and large angular momenta does not. As argued in \cite{Chen:2024qmn}, the criterion of allowable complex geometry can be modified so as to fit with the current gravity theory and our unphyisical geometry can be shown to satisfy the modified criterion.

One may still stick to the original criterion of allowable complex geometry as the conclusion is quite natural.
It might be possible to change the metric to be physical one after performing a proper gauge transformation. 
The decomposition by semi-classical saddles might not be unique and a nice decomposition by physical semi-classical saddles may exist. For instance, the four-point functions of heavy operators in a CFT$_2$ may be decomposed by conformal blocks in several ways, where the conformal blocks can be regarded as semi-classical saddles as discussed in \cite{Benjamin:2023uib}.

It is quite intriguing that the path integral of quantum gravity can be formulated even by integrating over unphysical geometries. One may think that this is a special feature of three-dimensional gravity and it does not occur in higher-dimensional case. Nevertheless, from the mini-superspace approach of quantum gravity, we can easily show that there exist saddle points which correspond to unphysical geometries, see, e.g., \cite{DiTucci:2020weq}. It is of course a different question whether such saddles are realized or not. We would like to examine this issue with the help of dual CFT as analyzed in this paper. Currently, there are many techniques available for non-perturbative aspects of quantum field theory, such as, conformal bootstrap, resurgent theory, supersymmetric localization. We hope to report on progresses on the issue in the near future.

\begin{acknowledgments}
We are grateful to Pawel Caputa, Shinji Hirano, Masazumi Honda, Hayato Kanno, Hikaru Kawai, Jun Nishimura and Tadashi Takayanagi for useful discussions.  The work is partially supported by Grant-in-Aid for Transformative Research Areas (A) ``Extreme Universe'' No.\,21H05187. The work of H.\,Y.\,C. is supported in part by Ministry of Science and Technology (MOST) through the grant 110-2112-M-002-006-. H.\, Y.\, C.\, would also like to thank Yukawa Institute for Theoretical Physics, Kyoto University for the hospitality when this work was being finished. The work of Y.\ H. is supported by JSPS Grant-in-Aid for Scientific Research (B) No. 23H01170. Y.\,T. is supported by Grant-in-Aid for JSPS Fellows No.\,22KJ1971. T.\,U. is supported by JSPS Grant-in-Aid for Early-Career Scientists No.\,22K14042.
\end{acknowledgments}

\bibliography{Resurgence}

\end{document}